\documentstyle[12pt]{article}
\begin{document}

\renewcommand{\baselinestretch}{1.5}
\newcommand\beq{\begin{equation}}
\newcommand\eeq{\end{equation}}
\newcommand\bea{\begin{eqnarray}}
\newcommand\eea{\end{eqnarray}}
\newcommand\co{correlations }
\newcommand\dy{dynamical }
\newcommand\om{\omega_q }

\centerline{\bf Harmonic Lattice Approximation and Density Correlations in}
\centerline{\bf the Calogero-Sutherland Model}
\vskip 1 true cm

\centerline{Diptiman Sen \footnote{Permanent Address: Centre for Theoretical 
Studies, Indian Institute of Science, Bangalore 560012, India} and R. K. 
Bhaduri}
\small
\centerline{\it Department of Physics and Astronomy, McMaster University,}
\centerline{\it Hamilton, Ontario L8S 4M1, Canada}
\vskip .25 true cm
\normalsize
\vskip 1.5 true cm

\noindent 
{\bf Abstract}
\vskip .5 true cm
For a one-dimensional model in which the two-body interactions are long-range
and {\it strong}, the system almost crystallizes. The harmonic modes of such
a lattice can be used to compute the ground state wave function and the 
dynamical density-density correlations. We use this method to calculate the 
density corrrelations in the Calogero-Sutherland model. We show 
numerically that the correlations obtained are quite accurate even if the 
coupling is not very large. Our method is considerably simpler than the ones 
used to derive the exact results, and yields expressions for the correlations 
which are easily plotted. Our equal-time correlations can be expanded in 
powers of the inverse coupling; we show that the two leading order terms agree 
with the exact results which are known for integer values of the coupling. 

\vskip 1 true cm
\noindent PACS numbers: ~63.20.-e, ~71.10.Pm

\newpage

\centerline{\bf I. INTRODUCTION}
\vskip .5 true cm

The Calogero-Sutherland-Moser model (CSM) [1-3] has received much attention 
recently due to its connection with a variety of interesting problems. Some 
examples are random matrix theory \cite{MEH}, quantum spin chains with 
long-range interactions \cite{HAL1,POL1}, Luttinger liquids \cite{HAL2}, 
Gaussian conformal field theories \cite{KAW}, edge states in a quantum Hall 
system \cite{YU}, generalized exclusion statistics [10-16], and nonlinear 
internal waves in a stratified fluid \cite{CHE}. 

Although the quantum mechanical energy spectrum of the CSM has been known for
a long time [1-3], its collective properties [18-22] and \dy density-density 
\co [23-28] are not yet expressed in an easily accessible form. The 
\co are best understood for three particular values of the coupling parameter 
$\lambda = 1/2$, $1$ and $2$ which are closely related to 
random matrix theory \cite{MUC,ZIR}. An impressive amount of information is
also available for all integer [25-27] and rational \cite{HA} values of
$\lambda$. However, even though these are all exact results, the mathematical
expressions are not very convenient for plotting, say, the space-time
dependence of the correlations, unless $\lambda$ (or its inverse) is equal to
a fairly small integer.

It therefore seems useful to pursue simple and approximate methods for 
obtaining the \co for general values of $\lambda$. In Sec. II, we present a
method for doing this which is actually exact in the limit of large $\lambda$.
In that limit, the particles almost freeze into a lattice \cite{POL1,SUT2}; 
the harmonic modes about such a lattice may then be used to compute the ground 
state energy, wave function and correlations. We use this idea to obtain the 
first two terms in an expansion in $1/\lambda$ for the ground state energy 
and the asymptotic expression for the equal-time density-density correlation; 
these agree with the known exact results \cite{FOR}. We show that the 
equal-time \co given by our method agree reasonably well with the exact result 
for all distances even if $\lambda$ is not very large. The \dy \co are also 
obtained with almost equal ease by our method, and are explicitly given for 
any space-time point. We discuss the low-temperature 
specific heat, and make a suggestion for improving the correlations by using a 
slightly modified phonon dispersion. In Sec. III, we briefly consider a more 
general model in which the two-body interaction decays as the power $\alpha$ 
of the distance. We find that the system behaves as a Luttinger liquid if 
$\alpha > 1$, and as a Wigner crystal if $\alpha < 1$; the Coulomb problem 
with $\alpha =1$ lies on the border between the two possibilities \cite{SCH}.

\vskip .5 true cm
\centerline{\bf II. CALOGERO-SUTHERLAND-MOSER MODEL}
\vskip .5 true cm

We will consider the form of the CSM in which particles on an infinite line 
interact pairwise through an inverse-square potential. The model may also be 
defined 
on a circle with periodic boundary condition \cite{SUT1}; the two versions 
of the model have identical physical properties in the thermodynamic limit 
in which the the number of particles $N$ and the length $L$ of the line (or 
circle) are simultaneously taken to infinity keeping the particle density 
$\rho_0 =N/L$ fixed. The Hamiltonian for particles on a line is given by
\beq
H ~=~ \sum_n ~\frac{p_n^2}{2m} ~+~ \frac{\hbar^2 \lambda (\lambda - 1)}{m}~ 
\sum_{l<n} ~\frac{1}{(x_l - x_n)^2} ~,
\label{hamcsm}
\eeq
where the dimensionless coupling $\lambda \ge 0$. To make the problem
well-defined quantum mechanically, we have to add the condition that
the wave functions go to zero as $\vert x_l - x_n \vert^{\lambda}$
whenever two particles $l$ and $n$ approach each other. For $\lambda = 0$
and $1$, the model describes free bosons and free fermions respectively. 
Since the two-body potential is singular enough to prevent particles from
crossing each other, we can choose the wave functions to be either symmetric 
(bosonic) or antisymmetric (fermionic). The energy spectrum is of course the 
same in the two descriptions. Two of the exactly known results for this model 
are as follows \cite{SUT1,SUT2}. In the thermodynamic limit, the ground state 
energy $E_0$ is given by
\beq
\frac{E_0}{N} ~=~ \frac {\pi^2 \hbar^2 \lambda^2 \rho_0^2}{6m} ~.
\label{e0}
\eeq
The sound velocity $v$ is given by the group velocity of long-wavelength
modes (wavelengths much bigger than the average particle spacing $1/\rho_0$),
\beq
v ~=~ \frac{\pi \hbar \lambda \rho_0}{m} ~.
\label{v}
\eeq

\vskip .5 true cm
\centerline{\bf A. Ground State Energy and Wave Function}
\vskip .5 true cm

Let us now consider the limit in which the coupling $\lambda \rightarrow 
\infty$. Then the potential energy term in (\ref{hamcsm}) dominates over the 
kinetic term. The potential energy is minimized if the particles sit at the 
sites of a lattice, so that the mean position of the $n$-th particle is 
$\langle x_n \rangle = n/\rho_0$ \cite{POL1,SUT2}. The total energy at this 
order in $\lambda$ is then given by 
\beq
\frac{E_0}{N} ~=~ \frac {\pi^2 \hbar^2 \lambda (\lambda -1) \rho_0^2}{6m} ~.
\label{e1}
\eeq
To the next order in $\lambda$, we have to consider the harmonic oscillations
about the mean positions. If we denote the fluctuation by $\eta_n = x_n -
n/\rho_0$, the Hamiltonian for the phonons is obtained by Taylor expanding 
the potential energy in (\ref{hamcsm}) to quadratic order. Thus
\beq
H_{ph} ~=~ \sum_n ~\frac{p_n^2}{2m} ~+~ \frac{3 \hbar^2 \lambda (\lambda -1) 
\rho_0^4}{m} ~\sum_{l<n} ~\frac{(\eta_l - \eta_n )^2}{(l - n)^4} ~,
\label{hamph}
\eeq
where we have ignored higher order anharmonic terms. We now
find the dispersion relation $\om$ for phonons with wavenumber $q$ to be
\bea
\om^2 ~&=&~ \frac{12 \hbar^2 \lambda (\lambda -1) \rho_0^4}{m^2} ~
\sum_{n=1}^{\infty} ~ \frac{1 - \cos (qn/\rho_0)}{n^4} ~, \nonumber \\ 
&=&~ \frac{\hbar^2 \lambda (\lambda -1)}{m^2} ~\Bigl(~\pi \rho_0 \vert q 
\vert ~-~ \frac{q^2}{2} ~\Bigr)^2 ~, \quad \vert q \vert \le \pi \rho_0 ~.
\label{disp}
\eea
(Note that $q$ can vary from $- \pi \rho_0$ to $\pi \rho_0$ in 
units of $2\pi /L$, so that the total number of modes is $N$). The sound 
velocity is given by $v = (\partial \om /\partial q)_{q=0} = \pi \hbar \rho_0 
\sqrt{\lambda (\lambda -1)}/m$. We see that this agrees with Eq. (\ref{v}) to
leading order in $\lambda$; we will therefore simply replace $\lambda (\lambda 
-1)$ by $\lambda^2$ in the expression (\ref{disp}) for $\om^2$. Namely, we 
use the exactly known sound velocity to correct the coefficient of our phonon 
dispersion. Henceforth we take
\beq
\om ~=~ \frac{\hbar \lambda}{m} ~\Bigl(~ \pi \rho_0 ~\vert q \vert ~-~ 
\frac{q^2}{2} ~\Bigr) ~.
\label{disp1}
\eeq
The zero point energy of the phonons is given by 
\beq
\sum_q ~\frac{\hbar \om}{2} ~=~ L ~\int_{-\pi \rho_0}^{\pi \rho_0} ~
\frac{dq}{2\pi} ~\frac{\hbar \om}{2} ~.
\eeq
When this is added to (\ref{e1}), we precisely recover Eq. (\ref{e0}); thus
the harmonic lattice approximation gives the ground state energy correctly
upto order $\lambda^2$ and $\lambda$. To get the next order term in the
energy, i.e. order $1$, we would have to use the original dispersion 
(\ref{disp}) as well as consider some of the anharmonic terms beyond $H_{ph}$;
we will not pursue this here.

We now use the harmonic approximation to write down the ground state
wave function following Ref. \cite{REA}. The unnormalized wave function for
a collection of decoupled phonons is given by
\bea
\Psi_0 ~&=&~ \exp ~\Bigl( ~- \frac{m}{2\hbar} ~\sum_q ~\om \eta_q \eta_{-q} ~
\Bigr) ~, 
\nonumber \\
\eta_q ~&=&~ \sum_{n=-\infty}^{\infty} ~\eta_n ~e^{iqn/\rho_0} ~.
\label{psi0}
\eea
Using the dispersion (\ref{disp1}), we find that 
\beq
\Psi_0 ~=~ \exp ~\Bigl( ~- \frac{\lambda \rho_0^2}{2} ~\sum_{l<n} ~
\frac{(\eta_l - \eta_n)^2}{(l-n)^2} ~\Bigr) ~.
\label{psi1}
\eeq
We see that this is equivalent to the exact ground state wave function of
the Hamiltonian (\ref{hamcsm})
\beq
\Psi_0 ~=~ \prod_{l<n} ~\vert \rho_0 (x_l - x_n) \vert^{\lambda} ~,
\eeq
if we Taylor expand
\beq
\log (\rho_0 \vert x_l - x_n \vert ) ~=~ \log ~\vert l - n \vert ~+~ 
\frac{\rho_0 (\eta_l - \eta_n )}{l-n} ~-~ \frac{\rho_0^2 (\eta_l 
- \eta_n )^2}{2 (l-n)^2} ~+~ \cdot \cdot \cdot ~.
\eeq

Before ending this subsection, let us write down the second quantized form for 
the Heisenberg operators $x_n (t)$,
\bea
x_n (t) ~&=&~ \frac{n}{\rho_0} ~+~ \int_{-\pi \rho_0}^{\pi \rho_0} ~\frac{dq}{2 
\pi} ~f_q ~\Bigl[ ~a_q ~e^{i (\frac{qn}{\rho_0} - \om t)} ~+~ a_q^{\dag} ~
e^{-i (\frac{qn}{\rho_0} - \om t)} ~\Bigr] ~, \nonumber \\
f_q ~&=&~ \Bigl( \frac{\hbar}{2m \rho_0 \om} \Bigr)^{1/2} ~.
\label{xnt}
\eea
Here 
\beq
[ ~a_q ~,~ a_{q^{\prime}}^{\dag} ~] ~=~ 2 \pi \delta (q - q^{\prime} ) ~.
\eeq
We can verify that $p_n (t) = m dx_n (t) /dt$ satisfies the equal-time 
commutation relation $[x_l , p_n ] = i \hbar \delta_{ln}$.

\vskip .5 true cm
\centerline{\bf B. Dynamical Correlation Function}
\vskip .5 true cm

We now use the harmonic lattice approximation to derive the \dy density-density
\co defined as
\bea
g(x,t) ~&=&~ \langle 0 \vert \rho (x,t) \rho (0,0) \vert 0 \rangle ~, 
\nonumber \\
\rho (x,t) ~&=&~ \sum_n ~\delta (x - x_n (t) ) ~,
\label{gxt}
\eea
where $x_n (t)$ is the Heisenberg operator given in (\ref{xnt}). To compute
this, we write the $\delta$-functions in (\ref{gxt}) as
\beq
\delta (x - x_n (t)) ~=~ \int_{-\infty}^{\infty} ~\frac{dq}{2\pi} ~e^{iq (x -
x_n (t))} ~.
\eeq
We then evaluate expectation values of the form $\langle 0 \vert \exp ~i(q x_n 
(t) - q^{\prime} x_l (0) ) \vert 0 \rangle$ using the Baker-Campbell-Hausdorff 
formula, 
\beq
e^{A+B} ~=~ e^A ~e^B ~e^{[B,A]/2} ~ 
\eeq
if $[B,A]$ commutes with both $A$ and $B$. Due to the logarithmic divergence 
at small momenta of an integral of the form $\int dk / \omega_k$, we find that 
this expectation value vanishes unless $q= q^{\prime}$. We finally obtain
\beq
g(x,t) ~=~ \rho_0 ~\int_{-\infty}^{\infty} ~\frac{dq}{2\pi} ~ 
\sum_{n=-\infty}^{\infty} ~\exp ~\Bigl[ ~iq(x - \frac{n}{\rho_0}) ~-~ 
\frac{q^2}{\rho_0^2} F(n,t) ~\Bigr] ~, 
\label{gxt1}
\eeq
where
\beq
F(n,t) ~=~ \int_0^{\pi \rho_0} ~\frac{dk}{\pi} ~\frac{\hbar \rho_0}{2 m 
\omega_k} ~\Bigl[ ~1 ~-~ e^{-i\omega_k t} ~\cos \Bigl( \frac{kn}{\rho_0} 
\Bigr) ~\Bigr] ~.
\label{fnt}
\eeq
It can be shown that the real part of $F(n,t)$ is always positive, unless both 
$n$ and $t$ are zero. We can therefore perform the $q$ integration in 
(\ref{gxt1}) to get
\beq
g(x,t) ~=~\rho_0^2 ~\sum_{n=-\infty}^{\infty} ~\Bigl( ~\frac{1}{4 \pi F(n,t)}~
\Bigr)^{1/2} ~\exp ~\Bigl[ ~- \frac{(x \rho_0 -n)^2}{4F(n,t)} ~\Bigr] ~.
\label{gxt2}
\eeq
The expansion in (\ref{gxt1}) can be understood in a physically intuitive
way as follows. Each lattice point $n$ contributes a Gaussian (correctly
normalized to unity) to $g(x,t)$; the squared width of the Gaussian is given
by the mean square fluctuation
\beq
\frac{2F(n,t)}{\rho_0^2} ~=~ \langle 0 \vert ~\Bigl( ~x_n (t) - x_0 (0) - 
\frac{n}{\rho_0} ~\Bigr)^2 ~\vert 0 \rangle ~.
\label{msf}
\eeq

Let us now study the equal-time \co $g(x,0) \equiv g(x)$ in some detail. From
(\ref{fnt}), we see that $F(n,0)$ is purely real, and $F(0,0)=0$. We therefore 
obtain
\beq
g(x) ~=~ \rho_0 ~\delta (x) ~+~ \rho_0^2 ~\sum_{n \ne 0} ~\Bigl( ~
\frac{1}{4\pi F(n,0)}~ \Bigr)^{1/2} ~\exp ~\Bigl[ ~- \frac{(x 
\rho_0 -n)^2}{4F(n,0)} ~\Bigr] ~,
\label{gx}
\eeq
where
\beq
\lambda F(n,0) ~=~ \frac{1}{2\pi^2} ~\int_0^{\pi} ~dy ~\frac{1 - \cos ~(yn)}{y~ 
-~ y^2/2\pi} ~=~ \frac{1}{2\pi^2} ~\int_0^{2 \pi} ~dy ~\frac{1 - \cos ~
(yn)}{y} ~.
\label{fn0}
\eeq
This gives us \cite{ABR}
\beq
\lambda F(n,0) ~=~ \frac{1}{2\pi^2} ~\Bigl[ ~\log ~(2 \pi e^{\gamma} n) ~-~
{\rm Ci} ~(2\pi n) ~\Bigr] ~,
\label{fn01}
\eeq
where $\gamma \simeq 0.57722$ is Euler's constant, and Ci denotes the cosine
integral. For large integer $n$, (\ref{fn01}) has an asymptotic expansion
beginning as
\beq
\lambda F(n,0) ~=~ \frac{1}{2\pi^2} ~\log ~(Cn) ~+~ \frac{1}{8\pi^4 n^2} ~+~
O \Bigl( \frac{1}{n^4} \Bigr) ~,
\label{fn02}
\eeq
where
\beq
C ~=~ 2 \pi e^{\gamma} ~\simeq ~11.191 ~. 
\eeq
The expansion in (\ref{fn02}) converges extremely rapidly; for $n=1$ and $2$, 
the difference between (\ref{fn01}) and the first two terms in (\ref{fn02}) is 
only about $0.1 \%$ and $0.01 \%$ respectively. For numerical purposes, 
therefore, we will use (\ref{fn02}) to compute $g(x)$ from (\ref{gx}).

We show a plot of $g(x)/\rho_0^2$ vs. $x\rho_0$ for $\lambda =2$ in Fig. 1; 
the solid line denotes the exact result \cite{SUT1}, while the dotted line 
shows our lattice approximation (\ref{gx}). The agreement appears to be 
reasonable even though $\lambda$ is not very large; for instance, the root mean
square fluctuation for nearest neighbor particles is given by (\ref{msf}) and 
(\ref{fn02}) to be $\sqrt{0.247 /\lambda} / \rho_0$, which is as large as $35 
\%$ of the lattice spacing $1/\rho_0$ for $\lambda =2$. In Fig. 2, we show the 
lattice approximation for $g(x)/\rho_0^2$ for $\lambda =4$. It is clear that 
the system shows an increasing tendency to crystallize as $\lambda$ increases.
We should note that the $\delta$-function at the origin has not been shown in 
Figs. 1 and 2; thus the $g(x)$ shown in the figures integrates to one hole,
\beq
\lim_{n \rightarrow \infty} ~2~ \int_{0}^{n/2} ~dx~ g(x) ~=~ \rho_0 ~(n - 1)~.
\eeq
We also note that in the lattice approximation, $g(x)$ does not quite vanish
at $x=0$ as it should in an exact calculation; however $g(0)$ does become small
rapidly as $\lambda$ increases.

We can Fourier transform $g(x)$ to obtain the static form factor 
\beq
S(q) ~=~ \int_{-\infty}^{\infty} ~dx ~e^{-iqx} ~g(x) ~.
\label{sq0}
\eeq
Since $g(x) \rightarrow 1$ for large $x$, $S(q)$ has a $\delta$-function at 
the origin. Hence it is easier to compute
\beq
S(q) ~-~ 2 \pi \rho_0^2 ~\delta (q) ~=~ \rho_0 ~+~ \rho_0^2 ~\int_0^{\infty} 
dx ~\cos ~(qx) ~\Bigl[ ~\sum_{n \ne 0} ~
\frac{\exp ~\Bigl[ -~ \frac{(x \rho_0 - n)^2}{4 F(n,0)} \Bigr]}{\sqrt{\pi
F(n,0)}} ~- ~2~ \Bigr] ~.
\label{sq}
\eeq
We show a plot of $S(q)/\rho_0$ vs. $q/\rho_0$ in Fig. 3 for $\lambda =2$ (the 
solid and dotted lines again denote the exact result and the lattice 
approximation respectively), and in Fig. 4 for $\lambda =4$. We will comment 
on the divergences at $q/\rho_0 = 2\pi$ in subsection C. (The small wiggles in 
Fig. 4 are due to numerical inaccuracies; the integral in (\ref{sq}) converges
very slowly if $\lambda$ is large. The power-law which governs the convergence 
will also be derived in the next subsection).

We now consider the \dy \co $g(x,t)$. It is convenient to introduce a 
dimensionless time variable 
\beq
{\tilde t} ~=~ v t \rho_0 ~.
\label{tt}
\eeq
Then
\beq
\lambda F(n,t) ~=~ \frac{1}{2\pi^2} ~\int_0^{\pi} ~dy ~\frac{1 ~-~ \cos ~(yn) ~
\exp ~\Bigl[ -i {\tilde t} (y ~-~ y^2 /2\pi ) \Bigr] }{y ~-~ y^2 /2 \pi} ~.
\label{fnt1}
\eeq
In the lattice approximation, plotting $g(x,t)$ is almost as easy as plotting
$g(x)$. In Figs. 5 and 6, we show $g(x,t)/ \rho_2$ vs. $x \rho_0$ for 
${\tilde t} =0.2$ and ${\tilde t} =2$ respectively, both for $\lambda =2$. 
The solid and dotted lines show the real and imaginary parts respectively. 
Note that the $\delta$-function at the origin which was not shown for $g(x)$ 
in Fig. 1 has now spread out and become visible at finite time in Figs. 5 and 
6; hence there is no hole in the real part of $g(x,t)$. For a small value of 
time $\tilde t$ and for $x$ close to the origin (Fig. 5), the $\delta$-function 
at the origin spreads out as a free particle with dispersion $\om = 
\hbar q^2 /2m$; this explains the large oscillations in both the real and 
imaginary parts. For a larger value of time (Fig. 6), and also for larger 
values of $x$ for any time, the particle feels the harmonic force of the 
lattice. Hence the behavior of $g(x,t)$ is different from the spreading of a 
free particle; in particular, the large oscillations get damped out. Finally, 
note that for ${\tilde t}= 2$, that $g(x,t)$ can differ appreciably from 
$g(x,0)$ only in the region $x \rho_0 = 0$ to $2$ since phonons can only 
propagate that far in that time. This is most strikingly visible in Fig. 5 in 
the imaginary part of $g(x,t)$ which vanishes rapidly beyond $x \rho_0 =2$.

Given $g(x,t)$, we can Fourier transform the $x$ coordinate to obtain a 
function which we denote by
\beq
S(q,t) ~=~ \rho_0 ~\sum_{n=-\infty}^{\infty} ~\exp ~\Bigl[ ~- ~
\frac{iqn}{\rho_0} ~-~ \frac{q^2}{\rho_0^2} F(n,t) ~\Bigr] ~.
\label{sqt}
\eeq
Thus the function $S(q)$ given in (\ref{sq0}) is equal to $S(q,t=0)$. If we 
now Fourier transform (\ref{sqt}) in time, we obtain the \dy structure factor
\beq
S(q,\omega ) ~=~ \int_{-\infty}^{\infty} ~dt ~e^{i\omega t} ~S(q,t) ~.
\label{sqo}
\eeq
This quantity can be written in terms of all the excited states $\vert \nu 
\rangle$ of the system,
\bea
S(q,\omega) ~&=&~ \rho_0 ~\sum_{\nu \ne 0} ~2\pi ~\delta (\omega - 
E_{\nu} + E_0 )~ \vert \langle \nu \vert \rho (q) \vert 0 \rangle \vert^2 ~, 
\nonumber \\
\rho (q) ~&=&~ \int ~dx ~\rho (x) ~e^{-iqx} ~=~ \sum_n ~e^{-iqx_n} ~.
\label{sqo1}
\eea

It is useful to consider the moments of $S(q,\omega )$ \cite{MUC} 
\beq
I_n (q) ~=~ \int_{-\infty}^{\infty} ~\frac{d\omega}{2\pi} ~\omega^n ~
S(q,\omega) ~.
\label{inq}
\eeq
These can be numerically obtained in a straightforward way from (\ref{sqt}), 
since $I_n (q) = i^n (\partial^n S(q,t) /\partial t^n)_{t=0}$. We have already 
considered $I_0 (q) = S(q)$ above. Since $F(0,0)=0$ and $(\partial F(n,t) /
\partial t)_{t=0} = i \delta_{n,0} ~\hbar \rho_0^2 /2m$, we find that the 
first moment
\beq
I_1 (q) ~=~ \frac{\hbar q^2 \rho_0}{2m}
\eeq
is independent of $\lambda$. This is an exact result following from the 
velocity independence of the two-body interactions, and it is called the
$f$-sum rule. The second moment $I_2 (q)$ is shown in Fig. 6 for $\lambda =2$;
we have actually plotted $I_2 (q) m^2/(\hbar^2 \rho_0^2 q^3)$ which is 
dimensionless and goes to a constant at $q=0$.

At this point, we observe the phenomenon of saturation by sound modes for
small momenta \cite{MUC}. Namely, for $q << 2 \pi \rho_0$, $S(q,\omega)$ is 
dominated by values of $\omega$ close to the phonon energy $\om$ given by 
(\ref{disp1}); thus $I_n (q)/ I_0 (q)$ approaches $\om^n$ as $q \rightarrow 
0$. Hence the $y$-coordinate in Fig. 6 should approach $\pi \lambda /2$ at 
$q=0$ which it does. For the same reason, the curves in Figs. 3 and 4 are 
linear near the origin with a slope of $1/(2 \pi \lambda)$, since $S(q) 
\rho_0$ approaches $\hbar q^2 /(2m \om)$.

The saturation by phonons at low energies also has a bearing on 
low-temperature thermodynamic properties of the CSM like the specific heat 
and pressure. We use the dispersion (\ref{disp1}) to compute the free energy 
per unit length $f$ taking the phonons to have zero chemical potential. Thus
\beq
\beta f ~=~ \int_{-\infty}^{\infty} ~\frac{dq}{2\pi} ~\ln ~(~1~-~ e^{-
\beta \hbar \om} ~) ~,
\eeq
where $\beta = 1/k_B T$. After evaluating this, we obtain the specific 
heat per unit length $C_V = -T \partial^2 f / \partial T^2$ to second order
in $T$. We find 
\bea
C_V ~&=&~ \frac{\pi k_B^2 T}{3 \hbar v} ~+~ \frac{6 \zeta (3) \lambda}{\pi}~ 
\frac{k_B^3 T^2}{\hbar m v^3} ~, \nonumber \\
\zeta (3) ~&=&~ \sum_{n=1}^{\infty} ~\frac{1}{n^3} ~.
\label{cv}
\eea
On comparing this with the exact result in Ref. \cite{ISA2}, we see that 
we have the correct coefficient for the term of order $T$; this is 
related to having the right phonon velocity at $q=0$. However we have a
coefficient $\lambda$ instead of $\lambda -1$ for the term of order $T^2$ in
(\ref{cv}). A different approach to the CSM based on collective field theory 
actually gives a dispersion \cite{SEN2}
\beq
\om ~=~ \frac{\hbar}{m} ~\Bigl( ~\pi \rho_0 \lambda ~\vert q \vert~-~ (\lambda 
-1) ~\frac{q^2}{2} ~\Bigr) ~,
\label{disp2}
\eeq
which leads to the correct coefficient $\lambda -1$ for the $T^2$ term. All
this suggests that we may get more accurate low-momentum or long-distance \co 
if we use the dispersion (\ref{disp2}) instead of (\ref{disp1}). Let us 
therefore use (\ref{disp2}) to compute $F(n,0)$ and then $g(x,0)$ following 
Eqs. (\ref{fnt}) and (\ref{gxt2}). The result is plotted and compared to the 
exact results for $\lambda=2$ in Fig. 8. On contrasting this with Fig. 1 which 
is based on the dispersion (\ref{disp1}), we see an improved agreement for 
large values of $x$, and, somewhat unexpectedly, for very small $x$ also. This
analysis again shows the close connection between the phonon dispersion and 
the \co in the lattice formalism.

\vskip .5 true cm
\centerline{\bf C. Large-$x$ Asymptotics for $g(x)$}
\vskip .5 true cm

We will now analytically evaluate $g(x)$ in (\ref{gx}) for large values of 
$x$, and show that the order $1$ and $1/\lambda$ terms agree with the exact 
results given in Ref. \cite{FOR}. We use the Poisson resummation formula
\beq
\sum_{n=-\infty}^{\infty} ~\phi (n) ~=~ \sum_{p=-\infty}^{\infty} ~
\int_{-\infty}^{\infty} ~dy ~e^{i2\pi yp} ~\phi (y) ~.
\label{poi}
\eeq
Given a function $\phi (n)$ defined for integer values, there are clearly many
ways to interpolate it to a function of a real variable $\phi (y)$. Eq. 
(\ref{poi}) says that different interpolations on the right hand side 
(assuming that the integrals and sum converge) must give the same answer for 
the left hand side. We will use this freedom to interpolate $F(n,0)$ to 
$F(y,0)$ taking the function in (\ref{fn02}), rather than the one in 
(\ref{fn01}). This is a convenient choice because the function in (\ref{fn02}) 
is smooth, unlike (\ref{fn01}) which contains oscillations. Thus, we use 
(\ref{poi}) with
\beq
\phi (y) ~=~ \Bigl( ~\frac{\lambda}{4 \pi f(y)} ~\Bigr)^{1/2} ~\exp ~\Bigl[ ~-~ 
\frac{\lambda (y-x\rho_0 )^2}{4 f(y)}~ \Bigr] ~, 
\label{phy}
\eeq
where $f(y)$ and its first two derivatives have the form
\bea
f(y) ~&=&~ \frac{1}{2\pi^2} ~\log ~(Cy) ~+~ \frac{1}{8\pi^4 y^2} ~+~ O
(\frac{1}{y^4}) ~, \nonumber \\
f^{\prime} (y) ~&=&~ \frac{1}{2\pi^2 y} ~-~ \frac{1}{4\pi^4 y^3} ~+~ O
(\frac{1}{y^5}) ~, \nonumber \\
f^{\prime \prime} (y) ~&=&~ -~ \frac{1}{2\pi^2 y^2} ~+~ \frac{3}{4\pi^4 y^4} ~
+~ O(\frac{1}{y^6}) ~.
\label{fy}
\eea

We now Taylor expand $\phi (y)$ around $y=x\rho_0$, and do the integrals over 
$y$ on the right hand side of (\ref{poi}). Upto order $1/\lambda$, this yields 
the general expression
\bea
\frac{g(x)}{\rho_0^2} = 1 + \frac{f^{\prime \prime} (x)}{\lambda} + 2 ~
\sum_{p=1}^{\infty} ~ \exp \Bigl( -\frac{4\pi^2 p^2 f(x)}{\lambda} \Bigr) & 
\Bigl[ & (1+ \frac{f^{\prime \prime} (x)}{\lambda} ) ~\cos (2\pi px) 
\nonumber \\
&& - ~\frac{4\pi p f^{\prime} (x)}{\lambda} ~\sin (2 \pi p x) ~\Bigr] . 
\nonumber \\
&&
\label{gx1}
\eea
We now use the explicit forms for $f(y)$ and its derivatives in Eq. (\ref{fy}) 
to obtain, upto order $1/x^2$,
\bea
\frac{g(x)}{\rho_0^2} = 1 - \frac{1}{2 \lambda \pi^2 x^2} + 2 ~
\sum_{p=1}^{\infty} ~(Cx e^{1/4 \pi^2 x^2})^{-2p^2/\lambda} & \Bigl[ & (1- 
\frac{1}{2 \lambda \pi^2 x^2} ) ~\cos (2\pi px) \nonumber \\
&& - ~\frac{2 p}{\lambda \pi x} ~\sin (2 \pi p x) ~\Bigr] . \nonumber \\
&&
\label{gx2}
\eea
(On the right hand sides of Eqs. (\ref{gx1}-\ref{gx2}), we have written $x$ 
instead of $x \rho_0$ for convenience). Eq. (\ref{gx2}) agrees with the 
asymptotic expression given in Ref. \cite{FOR} upto $O(\frac{1}{\lambda},
\frac{1}{x^2})$. We should make two comments here. First, the sum over $p$
runs upto infinity in (\ref{gx2}), while it only goes upto $p=\lambda$ in the
exact result for integer $\lambda$; however, the difference between the two
is exponentially small for large $\lambda$. Secondly, by going upto
order $1/\lambda^2$, we find terms involving $\log x$ which are not 
present in the exact result. This shows that the harmonic lattice approximation
fails at order $1/\lambda^2$, and that anharmonic terms must be included
to recover the correct asymptotics at that order. 

The oscillatory terms in (\ref{gx2}) imply that the 
Fourier transform $S(q)$ diverges at momenta $q=2\pi p \rho_0$ for
all nonzero integers $p$ satisfying $p^2 \le \lambda /2$. The divergence 
has a power-law form $\vert q - 2\pi p\rho_0 \vert^{(2p^2/\lambda)-1}$ if
$p^2 < \lambda /2$, and is logarithmic if $p^2 = \lambda /2$. We indeed
see a logarithmic divergence in Fig. 3 for $\lambda =2$, and a square-root
divergence in Fig. 4 for $\lambda =4$, both at $q = 2\pi \rho_0$.

The asymptotic analysis can be extended to the \dy \co $g(x,t)$. Let us define
\beq
s ~=~ \rho_0 ~\vert ~x^2 ~-~ v^2 t^2 ~\vert^{1/2} ~,
\eeq
and assume that $s >> 1$. Then we can show that the leading term in (\ref{fnt})
is given by
\beq
F(n,t) ~=~ \frac{1}{2\pi^2 \lambda} ~\log ~(Cs) ~.
\eeq
Following the above arguments, we find that
\beq
g(x,t) ~=~ \rho_0^2 ~\Bigl[ ~1 ~+~ 2~ \sum_{p=1}^{\infty} ~(Cs)^{-2p^2/
\lambda} ~\cos (2\pi x p \rho_0) ~\Bigr] 
\eeq
to leading order in the $1/\lambda$ expansion. 

The various power-law exponents derived above are in accordance with 
arguments based on the idea of Luttinger liquids \cite{HAL2} and conformal
field theory \cite{KAW}. In general, if the phonons in a one-dimensional
system have a low-momentum dispersion of the form $\om \simeq v q$, the
arguments presented here show that the prefactors of the oscillatory terms 
in (\ref{gx1}) must decay with the exponents $2\pi \hbar \rho_0 p^2/(mv)$.
In the next section, we will study the conditions under which a class of 
long-range interactions can lead to Luttinger liquid-like behavior.

\centerline{\bf III. OTHER LONG-RANGE MODELS}
\vskip .5 true cm

We consider a one-dimensional system in which the two-body interactions
decay as the power $\alpha$ of the distance. Using the lattice approximation,
we find that the long-distance \co are quite different depending on 
whether $\alpha > 1$, $\alpha =1$ (the Coulomb problem), or $0< \alpha < 1$
\cite{SCH}. This directly follows from the phonon dispersion which has the 
form
\beq
\om^2 ~\sim ~\sum_{n=1}^{\infty} ~\frac{1 - \cos (qn)}{n^{\alpha +2}} ~.
\eeq
(We set the density $\rho_0 =1$ in this section). For low momenta, $q << \pi$, 
we see that
\bea
\om ~&\sim&~ \vert q \vert \quad {\rm if} \quad \alpha > 1 ~, \nonumber \\
\om ~&\sim&~ \vert q \vert ~\sqrt{\log \vert q \vert} \quad {\rm if} \quad 
\alpha = 1 ~, \nonumber \\
\om ~&\sim&~ \vert q \vert^{(\alpha+1)/2} \quad {\rm if} \quad 0 < \alpha 
< 1 ~.
\eea
The asymptotics of the \co $g(x)$ are governed by the long-distance behavior
of 
\beq
F(n) ~\sim ~\int_0^{\pi} ~\frac{dq}{\om} ~\Bigl[ ~1 ~-~ \cos (qn) ~ \Bigr] ~.
\eeq
This diverges as $\log n$ if $\alpha > 1$ and as $\sqrt{\log n}$ if 
$\alpha =1$, but it does not diverge if $\alpha <1$. Correspondingly,
the prefactors of the oscillatory terms in $g(x)$ decay as powers of $x$
if $\alpha > 1$ (Luttinger liquid) and as exponentials of $-\sqrt{\log x}$
if $\alpha =1$, but they do not decay if $\alpha < 1$ (Wigner crystal)
\cite{SCH}. It is amusing that the Coulomb system, which is physically the 
most interesting of the three cases, lies exactly in between the Luttinger 
liquid and Wigner crystal scenarios.

\vskip .5 true cm
\centerline{\bf IV. DISCUSSION}
\vskip .5 true cm

We have seen that the harmonic lattice approximation is an useful 
technique from which many properties of the CSM can be derived readily without 
having to solve the Schr\"{o}dinger equation (\ref{hamcsm}) in great detail. 
We should however point out two problems which are either not clear or are 
beyond the reach of this approach.

\noindent
a) We have been unable to derive the one-particle Green's function, either 
bosonic or fermionic, from the phonon wave functions. This is because the 
sysmmtery (or antisymmetry) of the particle wave function does not enter the 
lattice formulation in a natural way; the lattice site and the particle labels 
are distinct from each other. This may be purely a technical difficulty.

\noindent
(b) It would be useful to systematically go to higher orders in the $1/ 
\lambda$ expansion by starting from the original phonon dispersion 
(\ref{disp}) and including the anharmonic terms perturbatively. This may 
provide a justification for replacing the dispersion (\ref{disp}) by either 
(\ref{disp1}) or (\ref{disp2}), and may enable us to extend our results to 
smaller values of $\lambda$.

\vskip .5 true cm

DS thanks F. D. M. Haldane for a stimulating discussion, and the Department of 
Physics and Astronomy, McMaster University for its hospitality during the 
course of this work. This research was supported by the Natural Sciences and 
Engineering Research Council of Canada.

\vskip .5 true cm
\noindent {\it Note Added:}

After writing this paper, we learnt that much of this work has been published 
earlier \cite{KRI,FOR2}; we thank P. J. Forrester for pointing this out. Ref. 
\cite{KRI} has gone much further by presenting expressions for $g(x,t)$ at 
finite temperature, for the singularities in $S(q,\omega)$, and for the 
equal-time Green's function.

\newpage

\noindent {\bf Figure Captions}
\vskip 1 true cm

\noindent 
{1.} Equal-time correlation $g(x)/\rho_0^2$ for $\lambda =2$. The solid and 
dotted lines denote the exact result and the lattice approximation 
respectively.

\noindent 
{2.} Correlation $g(x)/\rho_0^2$ for $\lambda =4$ obtained by the lattice 
approximation.

\noindent 
{3.} Static form factor $S(q)/\rho_0$ for $\lambda =2$. The solid and dotted 
lines denote the exact result and the lattice approximation respectively.

\noindent 
{4.} $S(q)/\rho_0$ for $\lambda =4$ obtained by the lattice approximation.

\noindent 
{6.} Dynamical correlation $g(x,t)/\rho_0^2$ for $\lambda =2$ and $vt \rho_0 = 
0.2$ obtained by the lattice approximation. The solid and dotted lines denote 
the real and imaginary parts respectively.

\noindent 
{6.} Correlation $g(x,t)/\rho_0^2$ for $\lambda =2$ and $vt \rho_0 = 2$ 
obtained by the lattice approximation. The solid and dotted lines denote the 
real and imaginary parts respectively.

\noindent
{7.} Second moment of \dy structure factor $I_2 (q) m^2 /(\hbar^2 \rho_0^2
q^3)$ for $\lambda = 2$ obtained by the lattice approximation.

\noindent 
{8.} Equal-time correlation $g(x)/\rho_0^2$ for $\lambda =2$. The solid and 
dotted lines denote the exact result and the lattice approximation using the
modified dispersion relation (\ref{disp2}) respectively.

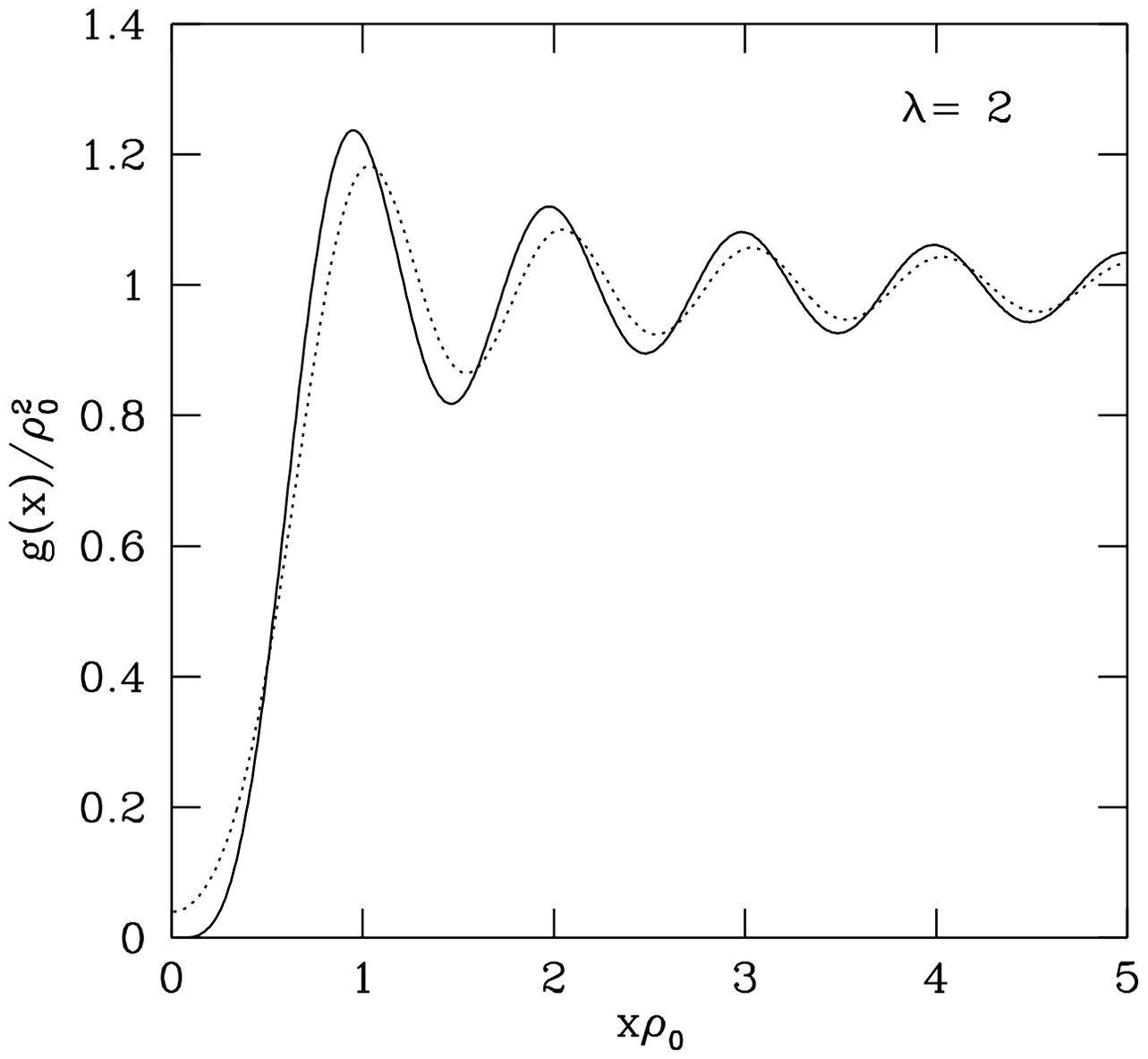
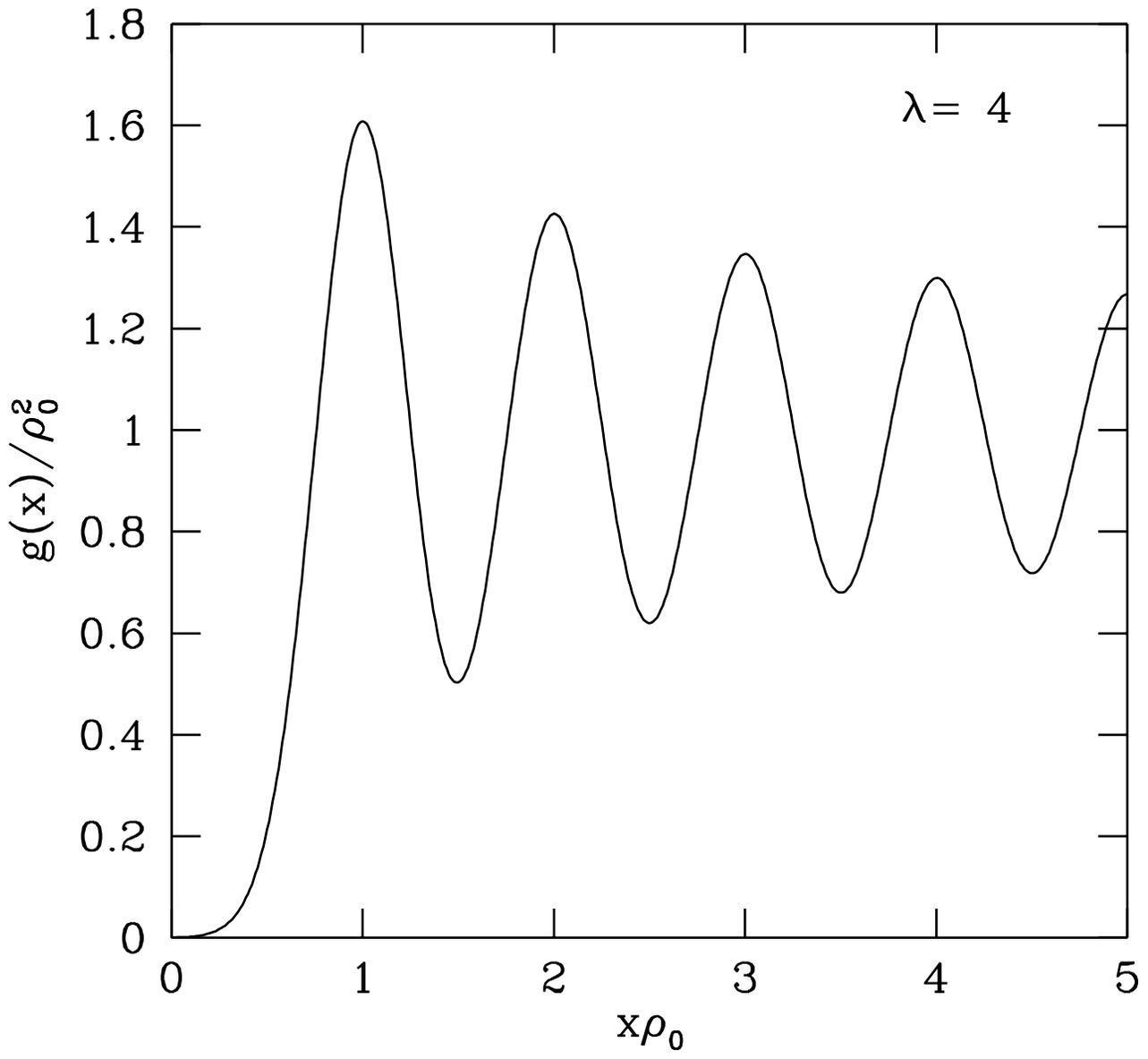
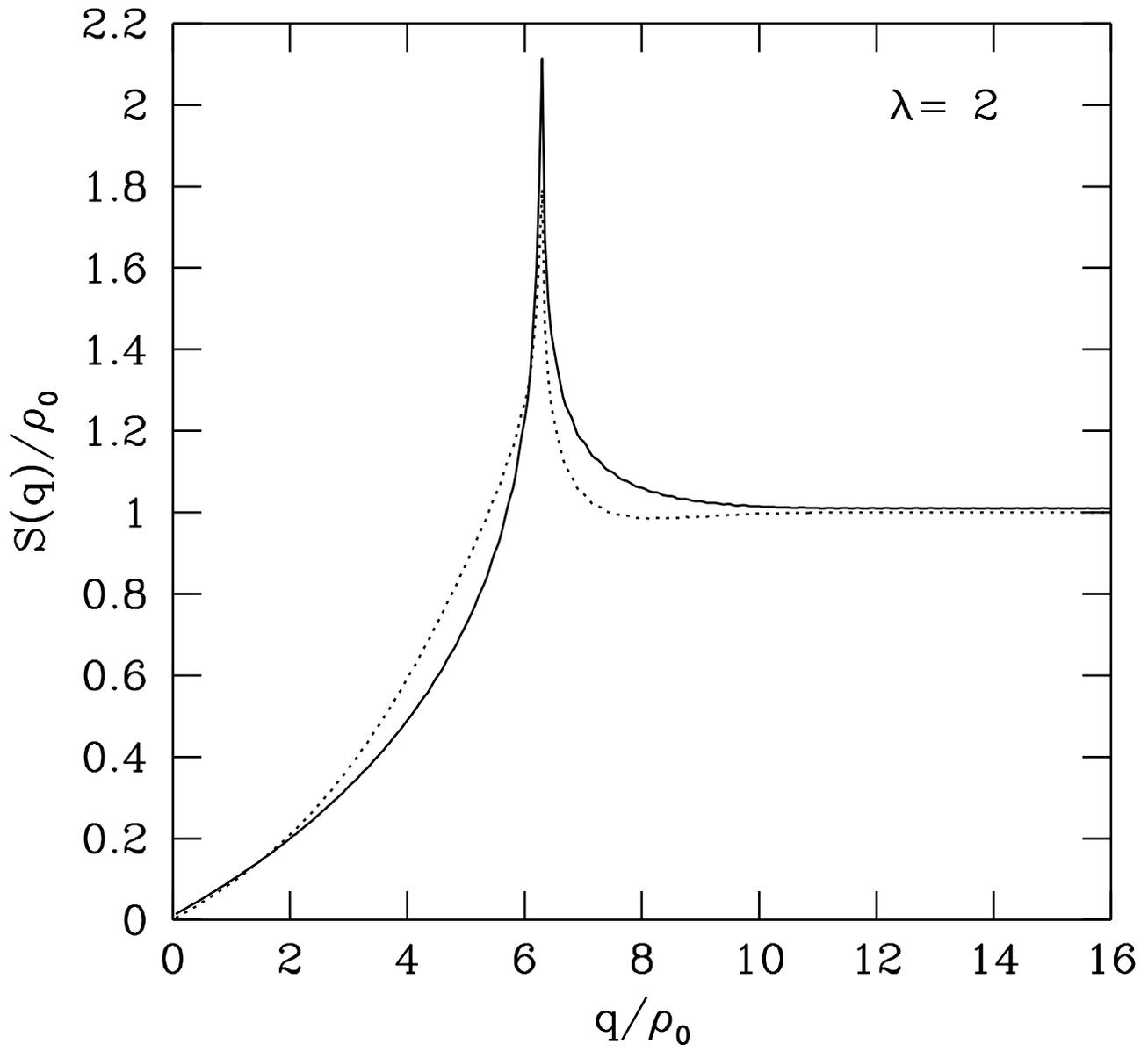
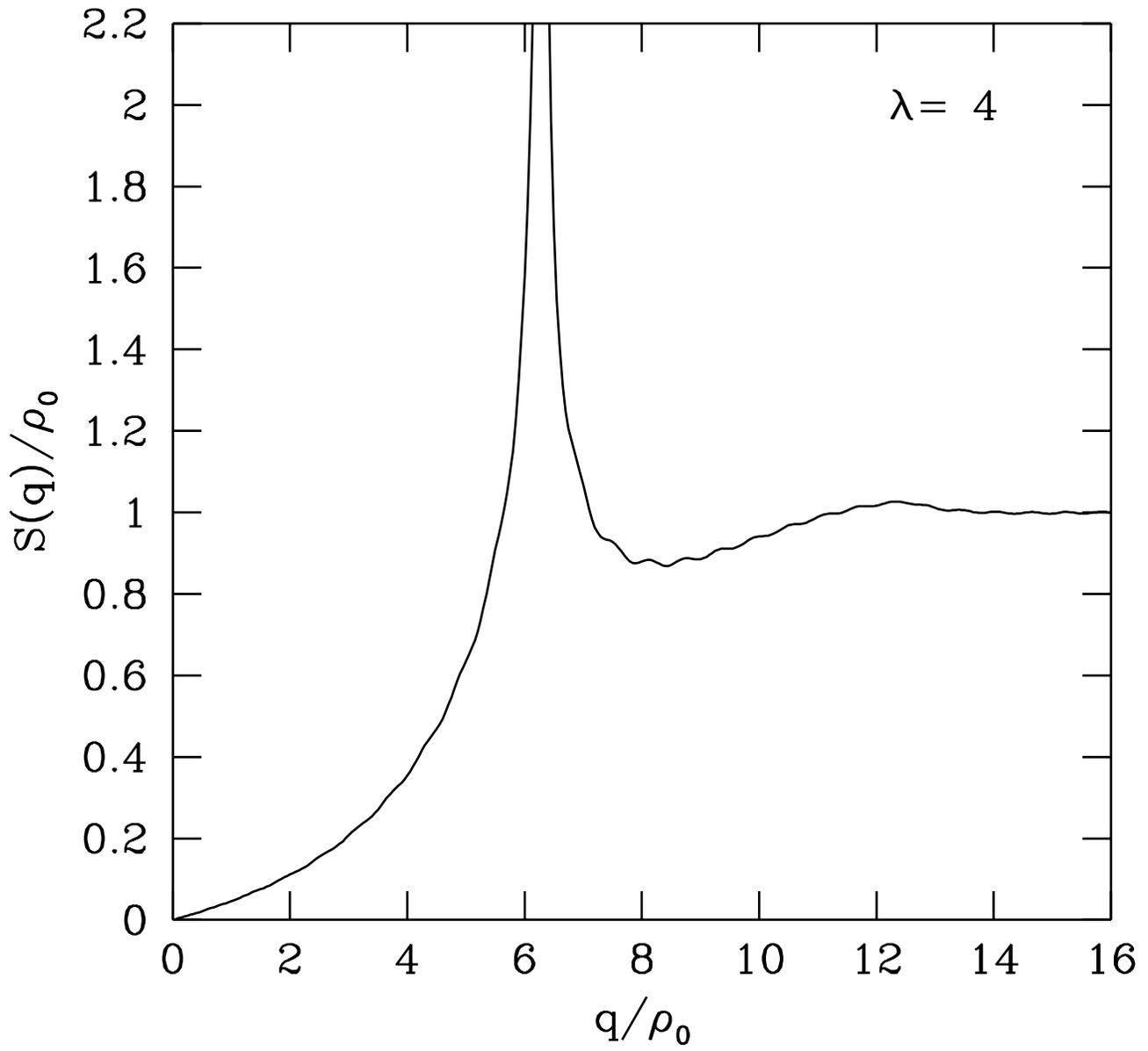
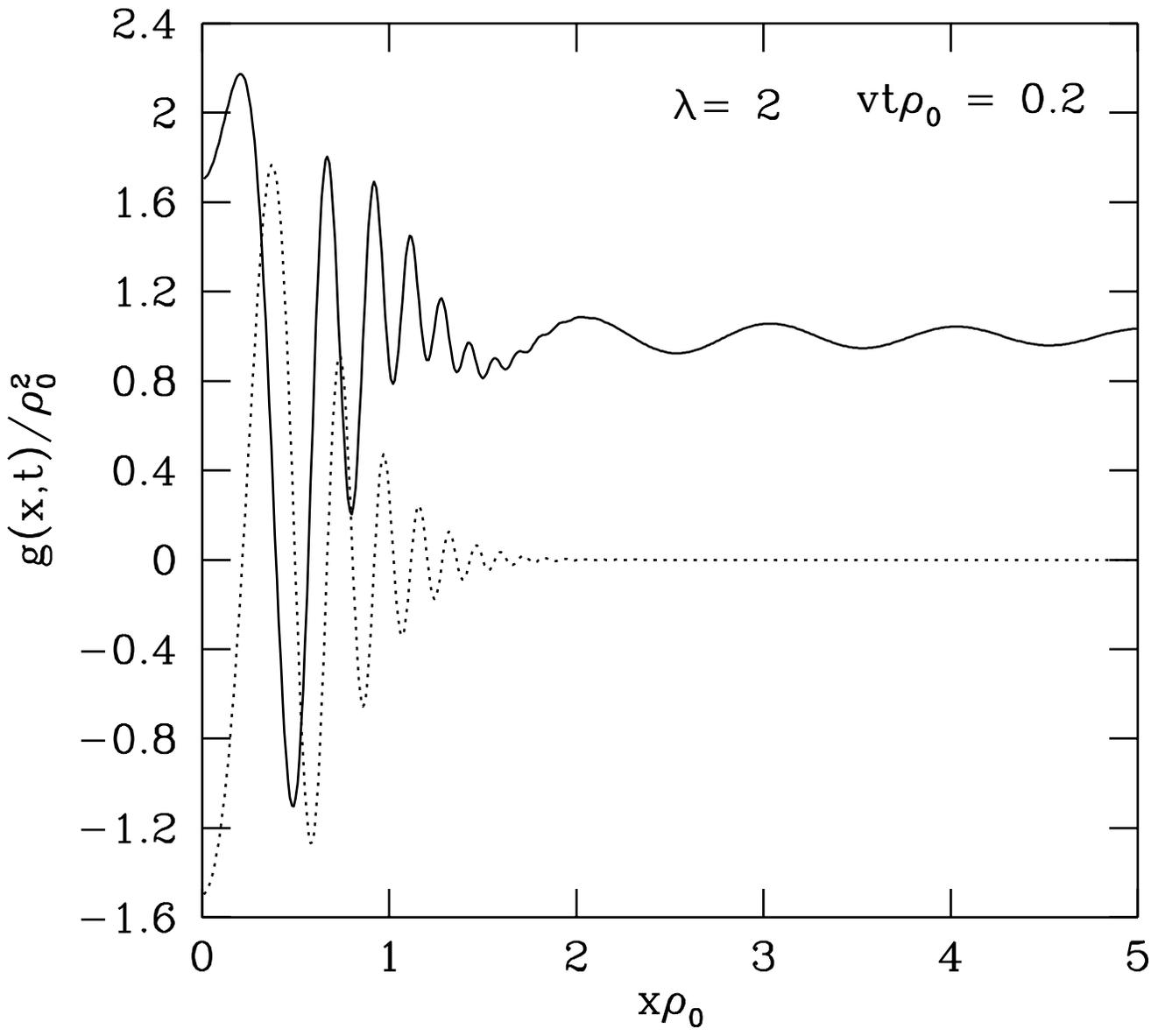
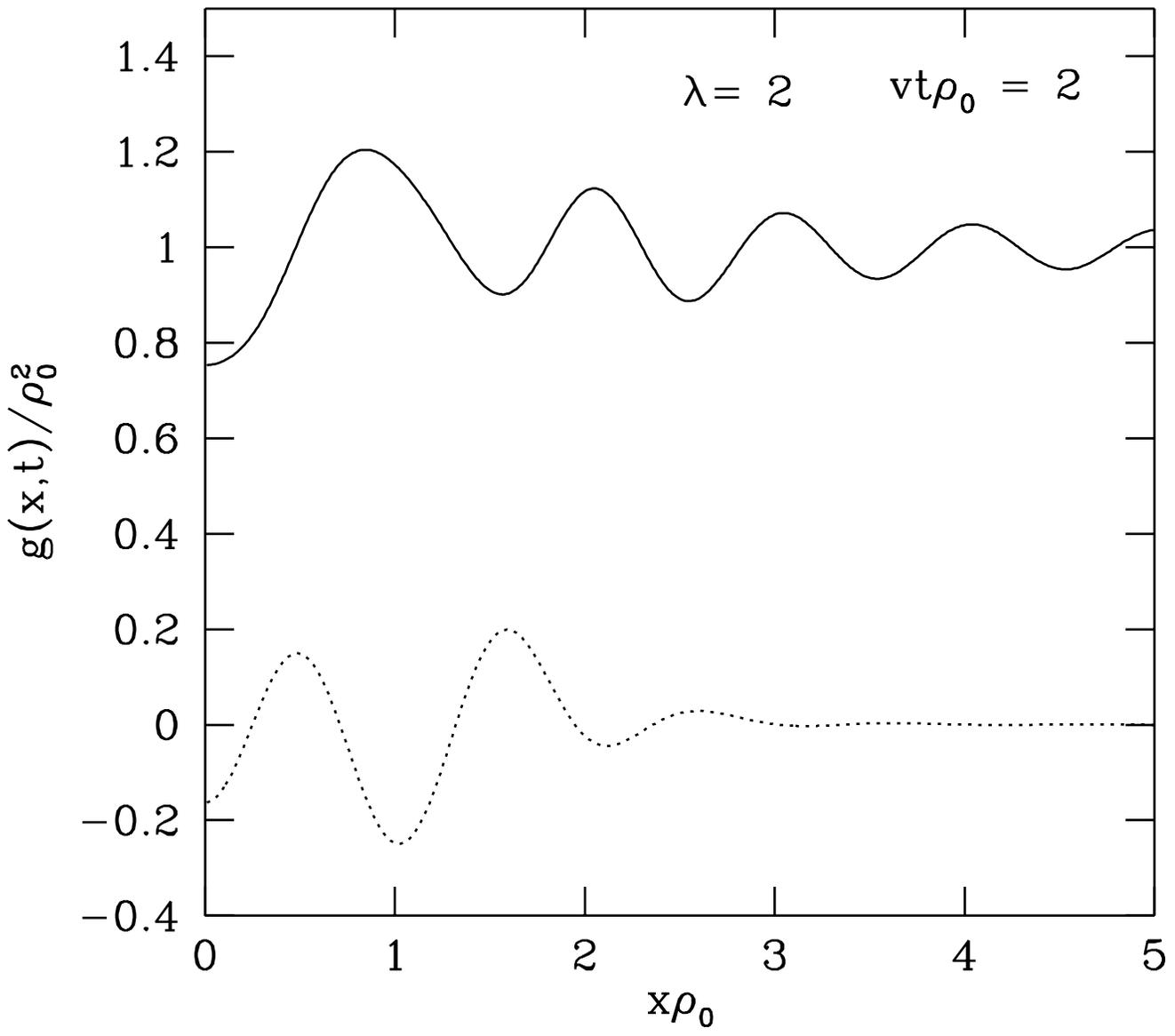
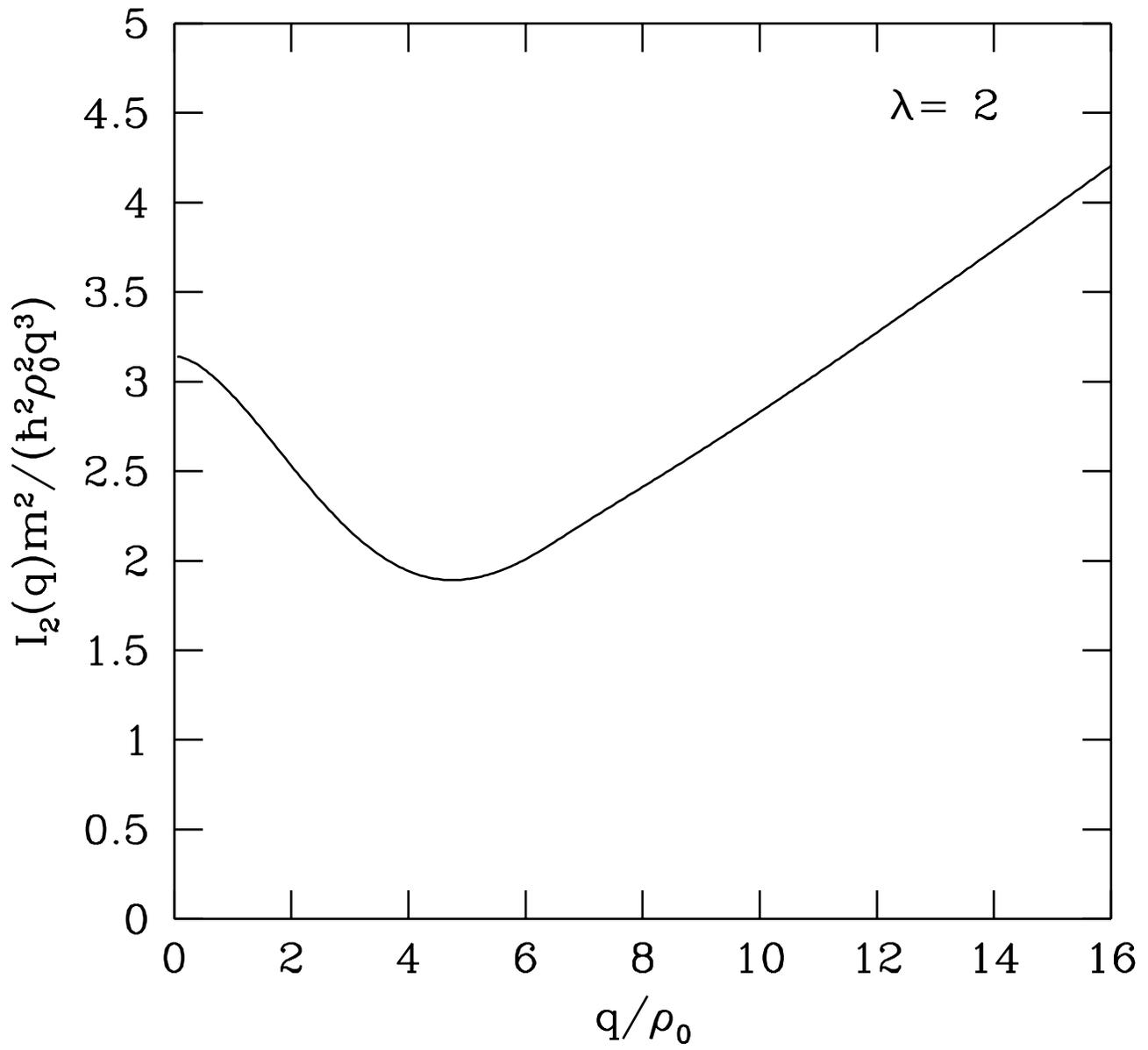
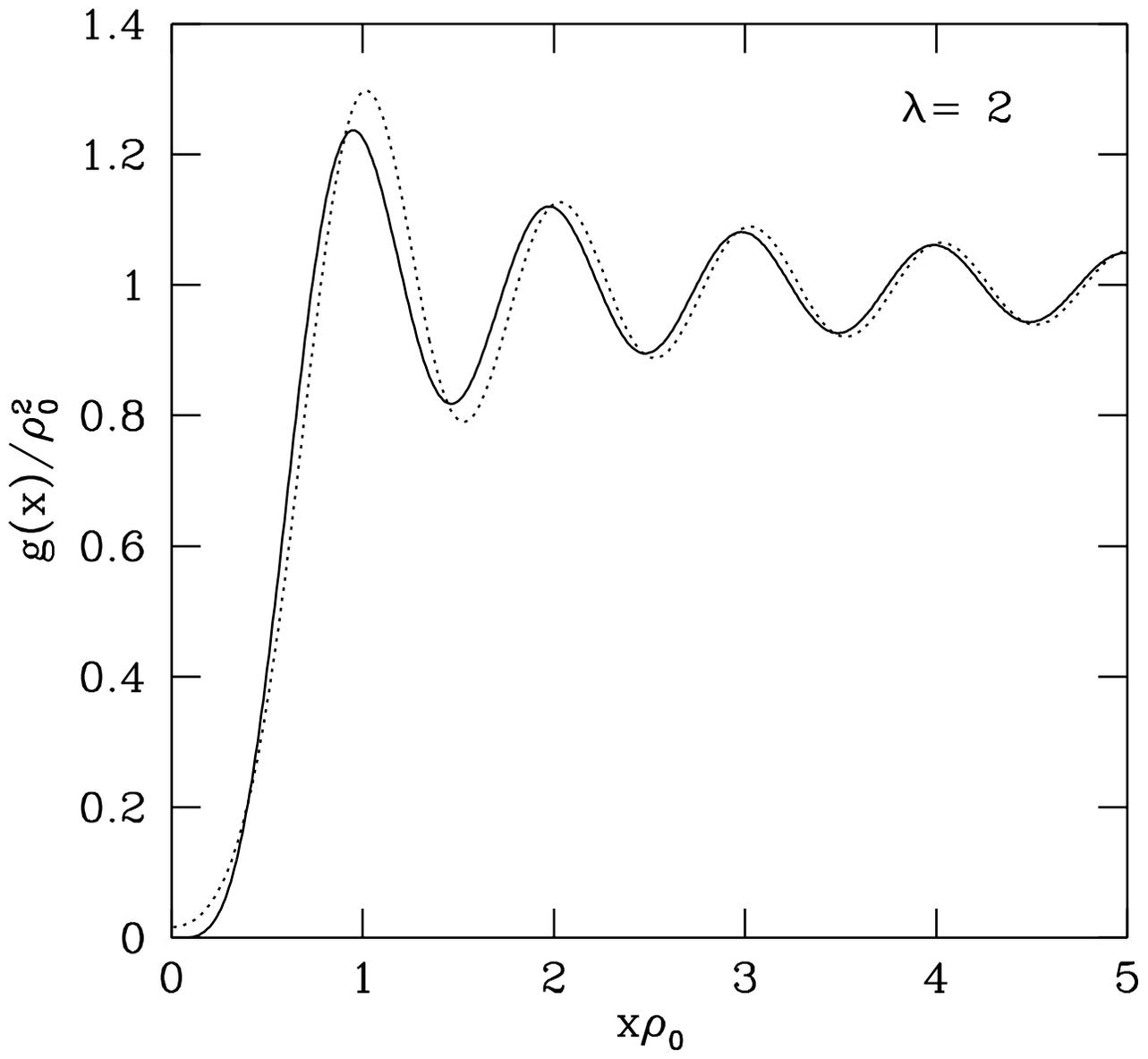
\end{document}